\voffset=.3in
\magnification=\magstep1                                                %
\def\ket{\rangle}
\def\bra{\langle}
\def\del{\partial}
\def\xx{|x\ket\bra x'|}
\def\kl{{\bf k}\lambda}
\def\kkll#1{{\bf k}_{#1}\lambda_{#1}}
\centerline{\bf Non-local Quantum Electrodynamics. II. Possibility of
correlated $2n$-photon absorption }
\centerline {\bf in gases leading to VERY High frequency
spontaneous emission}
\centerline {\bf and 
Very high order Harmonic Generation}
\vskip 2pc
\centerline {Krishna Rai Dastidar\footnote{\dag}
{Department of Spectroscopy. Electronic address :{\tt
spkrd@mahendra.iacs.res.in}} and T K Rai Dastidar\footnote *{this work is dedicated to his memory}} 
\centerline{\it Indian Association for the Cultivation of Science, Calcutta
700\thinspace 032, India}
\vskip 3pc
\centerline {\bf ABSTRACT}
\vskip 1pc
In a recent work [1] we expounded a non-local Quantum Electrodynamics (QED) 
which predicted a 
{\it linear two-photon absorption\/} by an atom placed in a laser field
of appropriate intensity and frequency. 
In this paper we extend our earlier work to show that the theory allows
for linear $2n$-photon absorption by gaseous matter where, under suitable 
conditions, $n$ may literally run upto thousands. The consequences of this 
extension of the theory are outlined and predictions are made which 
may be verified in laboratories.
\vfill\eject
\noindent{\bf I. INTRODUCTION}
\vskip 1pc
Recently [1] we showed that complex scalar fields remain covariant
under a non-local Abelian gauge transformation, and showed how a non-local
QED follows naturally from this gauge symmetry. (This paper will be referred
to as I in the following.) The result can of course
be trivially extended to spinor fields. From a practical point of view, the
most important result of this nonlocal QED is that under suitable conditions
that can be met in laboratory as well as cosmological environments, a {\it
two-photon absorption linear in intensity\/} can occur from radiation fields
travelling through gaseous matter.  The impact of this new QED on
atomic and molecular physics, particle physics and cosmology has been 
discussed in a series of papers [1-4]. In [2] we also suggested an
experimental test for this abovementioned result.

In paper I, it has been shown [1] that the correlated two-photon absorption 
in atoms can give rise to probability which is linear in laser intensity. 
In this paper we
 will show that this correlated two-photon absorption process
can be generalized to 2n-photon absorption process, where n can go upto 
thousands or more. Previously it has been shown that the correlated mode-pair 
of 
photons may exist under certain conditions, within the framework of non-local
Quantum Electrodynamics (QED). By second quantization of this non-local field, 
correlated two-photon annihilation and creation operators were developed and
it has been demonstrated that these operators operate on correlated photon number 
states by annihilating or creating two photons at a time respectively (one photon
from each mode). Hence one can get correlated vacuum field  in absence of any 
photons. Which means that the vacuum for the electromagnetic field is structured
by the presence of this new type of non-local correlated zero-photon mode pairs. 
Here we will generalize this vacuum field by considering 2n correlated modes, 
where n can go upto
a large number. But to make the zero point energy finite, one will have to put 
an upper limit to the value of n. The upper limit  for
the number n can be such that ${n\hbar\omega} \le {m_e} c^2$ where $m_e$ is the
rest mass of electron, c is the velocity of light and $\omega$ is the frequency of 
the photon.

It has been shown before that the correlated two-photon absorption can occur in
a single electron or two-electron (preferably correlated) system, under the 
conditions that (i) the time for two-photon absorption $\delta t << {1\over{\omega}}$,
where $\omega$ is the frequency of the photon and (ii) the number of photons 
(of energy flux I)
passing through atomic volume V i.e. ${VI \over{c\hbar\omega}}$ must be greater
than or equal to the number of outer-shell electrons ($n_e$) in the atom. 
Of course, above number of photons should be present within the volume V during
the time $\delta t$, which is much less than $1\over{\omega}$, so that the 
absorption can occur during this time interval. Therefore, there are three
parameters V, I and $\omega$ which can be controlled to satisfy the condition
$${VI\over {c\hbar{\omega^2}}} >> n_e \eqno(1)$$ 
For a single atom, V is fixed and if $\omega$ is given, one will have to increase
the laser intensity I, so that the number of photons exceeds the number of electrons 
present in the outer shell. On the other hand if V can be made large as in the 
case of cluster of atoms, the condition (1) can be easily satisfied for a lower 
laser intensity. In case of n-photon absorption, where n can be large on the 
order of 1000, the cluster will be a preferable candidate, since it can have 
radius one or two orders of magnitude larger than the single atom. Under this
condition absorption of 1000s' of photons will be more efficient for clusters
than for single atom for a given intensity of laser. Moreover, absorption of
1000s' of photons in a single atom, within the interval of time $\delta t$
becomes a more stringent condition than that in cluster of atoms where a large
number 
of electrons are available within the volume V.

In the previous paper it has been shown that the correlation among electrons
may affect the phenomenon of correlated two-photon absorption by two-electron
atoms. In this work we will not discuss about electron correlation, but it 
can be stated that the presence of electron correlation will also affect the
$2n$ photon correlated absorption process.

Within the framework of non-local QED, the vacuum field is structured due 
to the presence of 2n correlated zero-photon modes and hence the electrons 
excited by 2n correlated photon absorption can spontaneuosly emit a single
photon of energy $2n\hbar\omega_k$  due to the interaction with this 
correlated vacuum field. This is the phenomenon which can be attributed to the
emission of x-rays observed [7] in clusters when irradiated with
strong lasers of intensity $I \ge 10^{16} W/cm^2$.

Application of the above formalism in the area of matter-radiation
interaction has so far been confined to one- and two-electron atoms.
In the present paper it will be shown that the abovementioned
property of nonlocal gauge transformation symmetry can be extended in
such a manner as to be directly applicable to $n$-electron systems,
where $n$ can literally run upto thousands as for example in clusters. The
results of the interaction of such systems with laser radiation can,
according to this theory, be startling. One of the results that we are
going to prove is the possibility of $2n$-photon absorption linear in
intensity from a laser field, leading to the prospect of highly 
efficient $2n$-th order harmonic generation and/or spontaneous emission using 
gases consisting of
$n$-electron atoms, molecules or clusters. 
\vskip 1pc
\noindent{\bf II. THEORY}
\vskip 1pc
We begin by giving a brief summary of the work presented in I.

For a wavefunction (or classical field) $\phi(x)$ where $x\equiv ({\bf r},t)$, 
one of the ways to ensure the covariance (i.e. form-invariance) of the 
Lagrangian under an Abelian gauge transformation (GT) of the second kind 
$T=\exp(-ie\Lambda(x))$ [$\Lambda$ being any arbitrary function of 
{\it x\/}] is to replace the ordinary partial derivative
$\del_\mu\equiv \del/(\del x^\mu)$
in the Lagrangian by a {\it covariant derivative\/} $D_\mu$ involving a 
dynamical field [5]. 
$D_\mu$ is defined by the property that under the above
GT, $D_\mu\phi$ transforms in exactly the same manner as $\phi$ does~;
i.e. for infinitesimal $\Lambda, \delta(D_\mu\phi)=-ie\Lambda D_\mu\phi$.
For a charged particle one can define $D_\mu=\del_\mu+ieA_\mu(x)$, where 
$A_\mu=(A_0,-\bf A)$ is the electromagnetic field potential and $e$ plays the 
role of the {\it coupling constant\/} of the latter with $\phi(x)$. To ensure
covariance of $D_\mu$ under the above GT, the field potential $A_\mu$ should
transform as $A_\mu\to A_\mu+(\del_\mu\Lambda)$.
This, of course, agrees with the {\it gauge arbitrariness\/} of $A_\mu$
allowed by the {\it gauge-independence\/} of the electromagnetic field tensor 
$F^{\mu\nu}=\del^\mu A^\nu-\del^\nu A^\mu$.

Against this elementary background, we introduced in [1] a {\it non-local\/}
gauge transformation ${\cal T}=\exp(-ie\Lambda\xx)$, whose action 
on $\phi$ is defined by
$${\cal T}\phi\equiv\int\exp(-ie\Lambda(x,x'))\phi(x')\thinspace d^4x'
\eqno(1')$$
(A detailed discussion of this notation is given in [3], appendix A.)
For a charged particle, this allows the construction of a {\it new
covariant derivative} 
$${\cal D}_\mu={\del\over{\del x^\mu}}+{\del\over{\del x'^\mu}}+
ie{\cal A}_\mu\xx$$
where the action of ${\cal D}_\mu$ on $\phi$ is to be understood in the 
sense of eqn.\thinspace (1'), and where the nonlocal ${\cal A}$ transforms as
${\cal A}_\mu\xx\to {\cal A}_\mu\xx+\del_\mu\Lambda\xx+\del'_\mu\Lambda\xx$. 
The necessary and sufficient condition for gauge-independence of the 
corresponding nonlocal field tensor
$${\cal F}^{\mu\nu}\xx=(\del^\mu+\del'^\mu){\cal A}^\nu\xx -
(\del^\nu+\del'^\nu){\cal A}^\mu\xx\eqno(2)$$
is that the commutator $[\del^\mu,\del'^\nu]$ should vanish, i.e. the 
non-locality of the field should be of the EPR-type.

Under the action of this nonlocal field $\cal A$, a typical 
matrix element of an atomic transition $|i\ket\to |f\ket$ is given by 
$$M_{fi}(t)=-{e\over m}\bra\psi_f(x),n_f|{\bf p}.{\cal A}(x,x')|\psi_i(x'),
n_i\ket\eqno(3)$$
the integration running over $d^3${\bf r} and $d^4x'$. We have used the 
Coulomb gauge.

A Fourier expansion of this nonlocal potential can be carried out (see, e.g.
[6]) via a {\it double summation\/} over the usual photon modes~:
$${\cal A}(x,x')=\sum_{{\bf k}_1\lambda_1}\sum_{{\bf k}_2\lambda_2}
\left [C_{\kl}\hat\epsilon_{\kl}(\hat r,\hat r')\exp(i{\bf k}_1.{\bf r}
-i\omega_{{\bf k}_1}t+i{\bf k}_2.{\bf r}'-i\omega_{{\bf k}_2}t')+c.c.
\right ]\eqno(4)$$

Using standard methods (see any textbook on QED) the energy of this
nonlocal field in a volume $\Omega$ is given by
$$W_{2}={1\over 2}\sum_{\kl}(Y_{\kl}^2+4\omega_{\bf k}^2Z_{\kl}^2)\eqno(5)$$
where $Y_{\kl}=-i{\sqrt{\Omega/\pi}\over{2c}}2\omega_{\bf k}
(C_{\kl}-C^*_{\kl}), Z_{\kl}={\sqrt{\Omega/\pi}\over{2c}}
(C_{\kl}+C^*_{\kl})$. Each mode subscript ${\kl}$ here actually stands for a
{\it phase-correlated mode pair\/} $(\kkll{1},\kkll{2})$, and $2\omega_{\bf k}$
stands for $\omega_{{\bf k}_1}+\omega_{{\bf k}_2}$. The quantities $Y$ and $Z$
satisfy Hamilton's canonical equations of motion with $W_2$ as the 
hamiltonian, and in I we quantized the nonlocal QED by requiring that
$Y$ and $Z$ be $q$-numbers obeying the commutation relation
$$[Z_{\kl},Y_{{\bf k}'\lambda '}]=i\hbar\delta_{{\bf k}{\bf k}'}
\delta_{\lambda\lambda '}.$$
Ultimately, the nonlocal field energy comes out in terms of a {\it new pair
of creation and annihilation operators :}
$$W_{2}={1\over 2}\sum_{\kl}\hbar\omega_{\bf k}(b_{\kl}b^{\dag}_{\kl}+
b^{\dag}_{\kl}b_{\kl}),\quad {\rm where}\quad (b,b^{\dag})_{\kl}=
{1\over\sqrt{2\hbar\omega_{\bf k}}}(2\omega_{\bf k}Z_{\kl}\pm iY_{\kl})$$
Here $b$ and $b^{\dag}$ are {\it two-photon\/} annihilation and creation
operators, obeying the new commutation relations~:
$$[b_{\kl},b^{\dag}_{{\bf k}'\lambda '}]=2\delta_{{\bf k}{\bf k}'}
\delta_{\lambda\lambda '},\quad [b_{\kl},b_{{\bf k}'\lambda '}]=
[b^{\dag}_{\kl},b^{\dag}_{{\bf k}'\lambda '}]=0$$

A natural extension of the non-local GT (1) to the case of a two-particle
system would be the transformation ${\cal T}_2|x_1,x_2\ket\bra x'_1,x'_2|$
which, written out in full, becomes
$${\cal T}_2\phi(x_1,x_2)\equiv\int\int\exp(-ie\Lambda(x_1,x_2,x'_1,x'_2))
\phi(x'_1,x'_2)\thinspace d^4x'_1\thinspace d^4x'_2 \eqno(6)$$
This allows for a new covariant derivative
$${\cal D}_{2\mu}=\del_{1\mu}+\del_{2\mu}+\del'_{1\mu}+\del'_{2\mu}
+{\cal A}_{2\mu}|x_1,x_2\ket\bra x'_1,x'_2|\eqno(7)$$
whereupon the matrix element for the transition $|i\ket\to |f\ket$ becomes
$$M_{2_{fi}}(t)=-{e\over m}\bra\psi_f(x_1,x_2),n_f|({\bf p}_1+{\bf p}_2).
{\cal A}_2(x_1,x_2,x'_1,x'_2)|\psi_i(x'_1,x'_2),n_i\ket\eqno(8)$$

Let us now extend this formalism to correlated 2n-photon absorption in a 
n-electron system, by using the fourier expansion [6] of non-local vector 
potential as follows:

$${\cal A}(x_1,--,x_n,x'_1,--,x'_n)=\sum_{{\bf k}_1\lambda_1}--\sum_{{\bf k}'_n
\lambda'_n}
[C_{\kl}\hat\epsilon_{\kl}(\hat r_1,--,\hat r_n,\hat r'_1
,--\hat r'_n )$$
$$  \exp(i{\bf k}_1.{\bf r}_1+--+i{\bf k}_n
.{\bf r}_n
-i\omega_{{\bf k}_1}t_1--i\omega_{{\bf k}_n}t_n)$$
$$X \exp(i{\bf k}'_1.{\bf r}'_1+--+i{\bf k}'_n.{\bf r}'_n-
i\omega_{{\bf k}_1}' {t'}_1--i \omega_{{\bf k}_n}' {t'}_n)+ c.c ]\eqno(9)$$
Each mode subscript ${\kl}$ here actually stands for n
{\it phase-correlated mode pairs\/}\break
$({{\bf k}_1}\lambda_1, {{\bf k}'_1}{\lambda_1}'),--,({{\bf k}_n}\lambda_n,
{{\bf k}'_n}{\lambda_n}')$.

Hence the electric field amplitude ${\cal E}(x_1,--,x_n,x'_1,--,x'_n)$ can be obtained by evaluating
the time derivatives of {\cal A} with respect to times $t_1,--t'_n$.  
By using the usual method one can obtain the
field energy density in terms of new c-numbers $Y_{N{\kl}}$ and $Z_{N{\kl}}$
where $$Y_{N{\kl}}=-i{\sqrt{\Omega/\pi}\over{2c}}N\omega_{\bf k}
(C_{\kl}-C^*_{\kl})$$, $$Z_{N{\kl}}={\sqrt{\Omega/\pi}\over{2c}}
(C_{\kl}+C^*_{\kl})$$. 
 Here $N\omega_{\bf k}$
stands for $\omega_{{\bf k}_1}+--+\omega_{{\bf k}_n}'$. 
The quantities $Y$ and $Z$
satisfy Hamilton's canonical equations of motion with the total field energy
 as the 
hamiltonian. To second quantize the nonlocal QED,
$Y$ and $Z$ are considered to be $q$-numbers obeying the commutation relation
$$[Z_{N{\kl}},Y_{N{{\bf k}'\lambda '}}]=i\hbar\delta_{{\bf k}{\bf k}'}
\delta_{\lambda\lambda '}.$$
Ultimately, the field energy can be written in terms of a new pair of {\it
creation and annihilation operators} defined as 
$$(b_N,b^{\dag}_N)_{{\bf k}\lambda}={1\over{\sqrt{2\hbar\omega_k}}}
N\omega_{\bf k}(Z_{N{{\bf k}\lambda}}\pm iY_{N{\bf k}\lambda})\eqno (10)$$
these operators obey the commutation relations as follows
$$[(b_N)_{\kl},({b^{\dag}}_N)_{{\bf k}'\lambda '}]=N\delta_{{\bf k}{\bf k}'}
\delta_{\lambda\lambda '},\quad [(b_N)_{\kl},(b_N)_{{\bf k}'\lambda '}]=
[(b^{\dag}_N)_{\kl},(b^{\dag}_N)_{{\bf k}'\lambda '}]=0$$
These annihilation and creation operators operate on the correlated phopton 
number states as annihilating N number of photons or creating N muber of photons 
respectively as follows:

$${(b_N)}_{{\bf k}\lambda}\vert n_{{\bf k}\lambda}\rangle = {\sqrt n_{{\bf k}\lambda}}
\vert n_{{\bf k}\lambda} - N\rangle$$ 
and $$ ({b^{\dag}_N})_{{\bf k}\lambda} \vert n_{{\bf k}\lambda} \rangle =
\sqrt{n_{{\bf k}_\lambda}+N} \vert n_{{\bf k}\lambda} + N \rangle \eqno (11)$$

Therefore $(b_N)_{{\bf k}\lambda}$ and $({b^{\dag}}_N)_{{\bf k}\lambda}$ are the
N-photon annihilation and creation operators respectively. Hence the vector
potential and the field amplitude can be expressed in terms of these operators 
as shown in paper I.
Therefore the dipole transition matrix elements for correlated 2n-photon absorption
by n electrons (such as in clusters) can be given as:
$$ {\cal X}_{Nfi} = {\sqrt {2 \pi I}\over c}\langle\psi_{l_f,}({\bf r}_1,--{\bf r}_n)
\vert( {\bf r}_1+---+{\bf r}_n)$$
$$ . \hat \epsilon_{N\kl} ({\hat r}_1,----,{\hat r'}_n) \vert 
\psi_{l_i}({\bf r}'_1,--{\bf r}'_n) \eqno (12)$$
where the polarization vector $\hat\epsilon_{N{\kl}}$ can be expressed as
$$ \hat\epsilon_{\kl} ({\hat r}_1,--,{\hat r}_n,{{\hat r}'}_1,--,{{\hat r}'_n}) = 
\sum_{j} a_j \Pi_i ({\hat r}_i+{\hat r'}_i) P_j({\hat r}_i.{\hat r'}_i)$$
where j varies from 0 to $\infty$ and $i=1,2,---n$
Mod square of the matrix element ${\cal X}_{Nfi}$ will give rise to N-photon absorption
probability which varies linearly with laser intensity. One can get the selection
rule for the angular momentum by considering the angular integrals (as shown in paper I)
as $l_f=l_i$ and $l_f=l_i+N$
\vskip 2pc
\noindent {\bf III. DISCUSSIONS}
\vskip 2pc
It has been shown before, that the non-local theory has wide applications [1-4]
in the fields of atomic/molecular physics, cosmology and particle physics.
Recently it has been applied to explain the phenomenon of direct double 
ionization in atoms [8] in intense laser fields. Here it has been shown that
the non-local model of QED  can also be applied to explain the emission of
x-rays from cluster of atoms 
 when irradiated with strong laser field of intensity $> 10^{16} W/cm^2$.
In cluster of atoms the radius of the cluster can be 10 times larger than 
that of a single atom and hence the volume can be $10^3$ times larger than 
that of a single atom. It is obvious that the intensity (I) for correlated 
2n-photon excitation can be atleast three orders of magnitude less than that required
for the single atom (consider the relation (1)), in particular for a large number 
 (can be 1000 or more)  of photon absorption. Therefore the efficiency of
emission will increase with the increase in the size of the clusters.  But 
there is a limit 
in the increase of the size of clusters, since for efficient absorption the 
condition
(1) should be satisfied. Increase in the size of the cluster means addition of
 more atoms
and hence increase in the available number of electrons ($n_e$). Therefore if 
the increase 
in $n_e$ is faster than the increase in $V$, the condition (1) fails to be 
satisfied. Moreover
emission from clusters of high-Z atoms will be more effective, since 
in that case the available number of electrons will be more for the same size of the
the ciuster of low-Z atoms. 
Furthermore, there exists a cutoff intensity below which
x-ray emission from a cluster cannot be obtained (see quation (1)). 
For x-ray emission in a particular region of wavelength, the cutoff intensity will be 
different for  clusters of different atoms. 
According to this theory, since the 2n-photon correlated absorption 
should occur during a very short time $<< {1\over \omega}$, this 
phenomenon will
be more effective for short-pulsed lasers. Some of the above mentioned 
features have already been observed 
in recent experiments [7] on x-ray emission from clusters of different atoms 
in presence of
strong laser fields. Moreover in recent experiments [7] it has been found that 
together with the sharp higher order harmonic generations, x-ray emissions 
of relatively broad spectrum occur and persist for a long time. It has been 
shown above that once the atoms in cluster are excited by the
absorption of correlated 2n-photons, can emit spontaneously (photons of energy
 around $2n\hbar \omega$) due to the interaction of 
the excited atom/highly charged ions with the vacuum of correlated
modes. Since the duration of spontaneous emission depends on the natural
linewidth of the states from which it is emitted, it can be as long as few 
nanoseconds [7]. Let us consider a situation under which spontaneous emission 
of duration 1 nanosecond can occur from a highly excited atom/highly charged 
ion. If the duration of spontaneous emission of $12 ev$ photon can be of the order
 of 100 ns (e.g. spontaneous emission from $B^1\Sigma_u$ state to the ground state of
$H_2$ molecule), then for $1200 ev$ photon the duration of spontaneous emission  
can be of the order of 1 nanosecond, if the dipole transition moment from the ground
state of the atom to the highly excited state is less in magnitude by two 
orders 
from that of $H_2$ molecules (between the first excited state and the ground
state). Hence it is likely that the spontaneous emission from clusters can 
persist for 1 or two nanoseconds.
Another possibility is that the innershell ionization of atoms in clusters 
can occur
due to the highly efficient correlated 2n-photon absorption, followed by x-ray 
emission. 

The wavelength dependence of x-ray emission from cluster
 has been observed to be more efficient for shorter wavelength
lasers (248 nm) than that for the longer one (800 nm). This may be due to the fact that
for long wavelength lasers the number of photons required for a particular
transition is much larger than
that for short wavelength lasers. Therefore it will lead to
 higher order transitions
and hence correlation may be destroyed during the process of absorption.    

Most striking feature of this type of absorption in the non-local picture of the
electromagnetic field is that in this correlated 2n-photon
absorption process, the probability for absorption varies linearly with laser 
intensity, thus making the absorption process much more efficient than that for
multiphoton transition in the local picture of the electromagnetic field. 
Therefore within the framwework of non-local QED, the 2n-photon correlated 
absorption and hence the spontaneous emission/ higher order harmonic 
generation from the cluster
will be very much  effective in presence of very strong laser 
($ I > 10^{16} W/cm^2$) field.

Moreover in the field of interaction of matter with radiation, one can
predict that if the 2n-photon correlated absorption of
radiation by the electrons in matter is feasible (in this model of non-local QED),
leading to processes such as pair production, the phenomenon of pair 
production will be visible for much lower intensity than that required for 
conventional transitions.

In conclusion, it has been shown that the two-photon correlated absoption 
model in the 
non-local QED can be extended to any arbitrary 2n-photon correlated absorption 
and hence leading to efficient harmonic generation and/or spontaneous emission
due to the interaction with structured vacuum ( of correlated
 modes). This type of absorption can occur in cluster of atoms when irradiated
with short-pulsed intense radiation field and the efficiency of this process 
will be much greater than the conventional multiphoton processes, 
since the 2n-photon absorption in the non-local model is linearly dependent 
on laser intensity. With this model of non-local QED, some experimentally
observed features could be explained and some predictions have been made 
which can be verified in the laboratory.

\vskip 2pc
\noindent {\bf ACKNOWLEDGEMENT}
\vskip 2pc 
K.R.D. is thankful to Prof. J. P. Connerade for helpful discussions.

\vfill\eject

\noindent{\bf REFERENCES}
\vskip 1pc
\item{[1]} T K Rai Dastidar and K Rai Dastidar, Mod.Phys.Lett. A{\bf 13},
1265 (1998), Errata A{\bf 13}, 2265 (1998)~; hep-th/9902020
\item{[2]} T K Rai Dastidar, Mod.Phys.Lett. A{\bf 14}, 1193 (1999) ;
quant-ph/9903043
\item{[3]} T K Rai Dastidar, Mod.Phys.Lett. A{\bf 14}, 2499 (1999) ;
quant-ph/9903053
\item{[4]} T K Rai Dastidar and K Rai Dastidar, Mod.Phys.Lett. A{\bf 14},
2555 (1999)~; hep-th/9906133
K. Rai Dastidar and T.K. Rai Dastidar, India J. Phys. (in press).
\item{[5]} Introduction of a dynamical field is a sufficient but not a 
necessary condition for constructing a covariant derivative~; a {\it pure
gauge field\/}, which may be taken to the limit $0^+$, can be used as well. 
See T K Rai Dastidar and K Rai Dastidar,
Nuovo Cimento {\bf 109B}, 1115 (1994)~; Mod. Phys. Lett. A{\bf 10}, 1843
(1995). The most important finding of these papers was that, contrary to
conventional wisdom, {\it free-particle equations of motion\/} in both
non-relativistic as well as relativistic Quantum Mechanics are covariant
under Abelian gauge transformations of the second kind.
\item{[6]} I M Gelfand and G E Shilov, {\it Generalised Functions\/}
(Academic Press, 1964) Vol.~I, Chap.~2, Eq. (1.3)(1)  
\item{[7]} T. Ditmire, T. Donnelly, A.M. Rubenchik, R.W. Falcon and M.D. Perry,
Phys. Rev. A, {\bf 53}, 3379 (1996); H. Honda, E. Miura, K. Karsura, E. Takabashi
and K. Kondo, Phys. Rev. A, {\bf 61}, 023201 (2000)
\item{[8]} K. Rai Dastidar (communicated)
\bye